\documentclass{elsart}

\usepackage{graphicx}

\usepackage{amssymb}

\begin{document}

\begin{frontmatter}
\title{Transverse Kinematics of Ions Stored in an Electrostatic Ion Beam Trap }
\author[1,2]{D. Attia}
\author[1]{D. Strasser}
\author[1]{O. Heber}
\author[3]{M. L. Rappaport}
\author[1]{D. Zajfman}

\address[1]{Department of Particle Physics,
Weizmann Institute of Science, Rehovot 76100, Israel}
\address[2]{Laboratoire Kastler Brossel,
\'Ecole Normale Sup{\'e}rieure et Universit{\'e} Pierre et Marie
Curie, Boite 74, 4 Place Jussieu, F-75252 Paris CEDEX 05, France}
\address[3]{Physics Services,
Weizmann Institute of Science, Rehovot 76100, Israel}

\begin{abstract}
We present experimental results, as well as numerical simulations,
for the transverse velocity distribution of ions stored in an
electrostatic ion beam trap. The measurements indicate that the
transverse velocity spread is about 1\% of the longitudinal
velocity, and that the ions fill the whole transverse stable phase
space. We also demonstrate that ion losses from the trap due to
multiple scattering with molecules from the residual gas is an
important factor limiting the lifetime of the beam.

\end{abstract}


\end{frontmatter}

\section{Introduction}

The use and development of ion trapping techniques, which started
about 50 years ago \cite{cite1}, have led to a broad range of
discoveries and new experiments in physics and chemistry. In
particular, one can cite high precision spectroscopy, mass
measurements, particle dynamics, nuclear and atomic processes and
the measurement of fundamental constants \cite{ionTrapping}.
During the last few years, a new type of ion trap has been
developed in which ion beams, instead of ion clouds, are
trapped\cite{EIBT1,EIBT2}.  This new trap stores particles using
only electrostatic fields and works on a principle similar to that
of an optical resonator. The main advantages of the trap are the
possibility to trap fast (keV) beams without need of deceleration,
the well defined beam direction, easy access to the trapped beam
by various probes, and simple requirements in terms of external
beam injection. Different types of experiments have already been
performed with these traps, such as the measurement of metastable
state lifetimes of atomic and molecular ions \cite{meta1,meta2},
the lifetimes of metastable negative ions \cite{life1,life2}, and
electron impact detachment cross sections of negative clusters
\cite{adi}. Cluster cooling has also been observed \cite{yoni}.
Interesting dynamics of the ion motion have been discovered, such
as self-bunching (due to the so-called negative mass instability
phenomenon) and the possibility of using simple phase space
manipulation to reduce the velocity spread \cite{bunch1,sarah}.
Electrostatic ion storage rings\cite{Moller:97,Moller:01} have
also been used during the last several years in a variety of
experiments\cite{He-:01,Hansen:01}.

Although the motion of the ions in the trap can be readily
simulated, no measurements of the transverse velocity
distribution (TVD) of the stored beam have hitherto been performed. The TVD is needed to
understand the trapping efficiency, as well as the beam loss
processes, especially the ones related to multiple scattering. We describe here the method
that we have developed to
characterize the TVD of the stored
ions. The results are compared to numerical trajectory simulations,
which confirm that multiple scattering is the dominant loss
process in these traps, and that the available area of the stable
transverse phase space directly influences the lifetime of the trapped ion
beam.

\section{Experimental setup}

\subsection{Ion trap}

\begin{figure}[h]
\begin{center}
\includegraphics[width=16.cm]{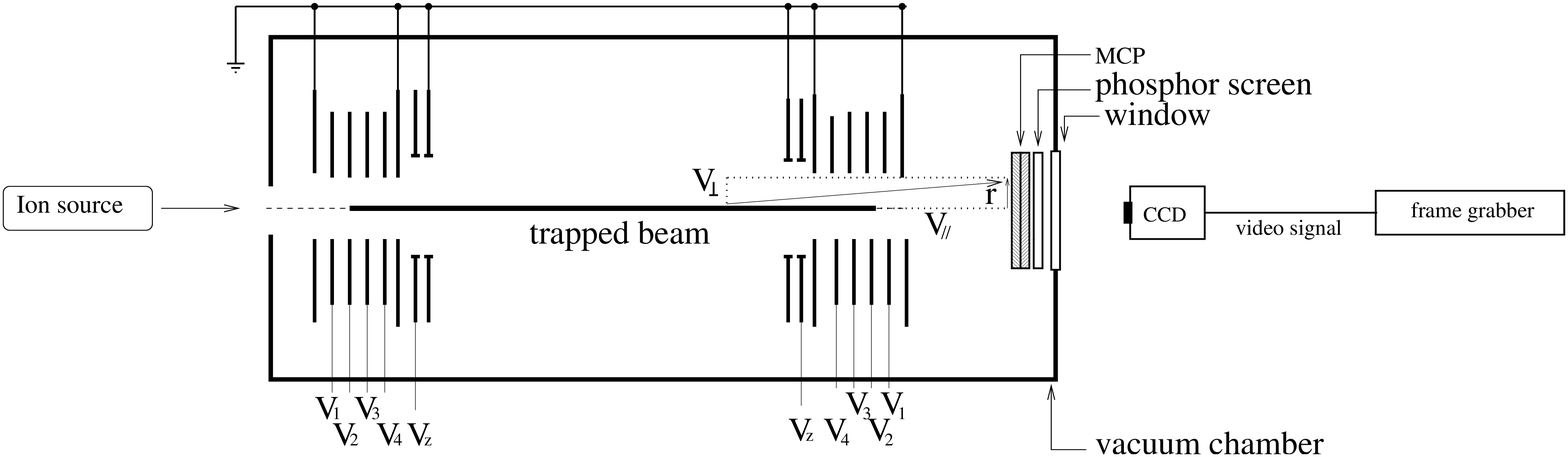}
\end{center}
\caption{}
\label{expsetup}
\end{figure}

Figure \ref{expsetup} shows a schematic view of the
electrostatic ion trap and the detection system. Two different
setups were used for creating the ions. For light ions,  an
electron impact ionization source was used and the ions were mass
selected with two magnets. For heavier species, a
matrix assisted laser desorption and ionization
(MALDI)\cite{maldi} source was used to create an ion
bunch, which was mass selected using time of flight. In
both cases, the ions were accelerated to an energy of 4.2 keV.
Three different types of ions were used in this work: Au$^+$
(m=197) and singly charged angiotensin II (m=1046) (both
produced by the MALDI source), and Ar$^+$ (m=40) (produced by the
electron impact source). After focusing and collimation, the beam
is directed into the ion trap along its axis. A complete
description of the ion trap is given in Ref. \cite{EIBT2}, and only
the details relevant for the present experiment will be given
here.

The trap is made of two identical cylindrically symmetric
``electrostatic mirrors'' that both trap the beam in the
longitudinal direction and focus it in the lateral direction. Upon
injection, the entrance set of electrodes (left side in
Fig.~\ref{expsetup}) is grounded so that the ion bunch can reach
the exit mirror (right hand side in Fig.~\ref{expsetup}), where
they are reflected. Before the reflected bunch reaches the
entrance electrodes, the potentials of these electrodes are
rapidly switched on ($\sim$ 100 ns rise time) to the same values
as those of the exit electrodes. For proper choices of voltages,
the ions bounce back and forth between the two mirrors, their
lifetime being limited mainly by collisions with the residual gas
molecules. The low pressure in the trap, of the order of
5$\times$10$^{-10}$  Torr when the electron impact source was
used, and 4$\times$10$^{-11}$ Torr for the MALDI setup, is
maintained by a cryopump.

Each electrostatic mirror comprises eight electrodes. The
potentials of the electrodes labeled $V_1$ to $V_4$ and $V_z$ in Fig.
\ref{expsetup} are independently adjustable. The other
electrodes are always grounded. Thus the 228 mm long central region of the
trap between the two innermost electrodes is essentially field-free.
The diameter of the central hole is
16 mm in the outer six electrodes and 26 mm in the two innermost
electrodes. The distance between the outermost electrodes is 407
mm.

In order for the ions to be trapped, the electrode potentials have
to satisfy certain conditions. It is well known that many
principles of geometric optics can be applied to ion optics. In
fact, our trap is based on a optical resonator made of two
cylindrically symmetric mirrors \cite{Pedersen1}. For an optical
resonator with identical mirrors and a Gaussian beam, the
stability criterion (for a beam close to the symmetry axis) is
related to the focusing properties of the mirrors:
\begin{equation}
L/4 \leq f \leq \infty, \label{eq:stable}
\end{equation}

where $f$ is the focal length of each mirror and $L$ is the
distance between them. This condition is easy to fulfill with the
above design. Another obvious requirement is that the maximum
potential on the mirror axis, $V_{max}$,  has to be high enough to
reflect the ions , i.e., $qV_{max} > E_k$, where $q$ is the charge
of the ions and $E_k$ is their kinetic energy.

Some important aspects of the design should be emphasized.
First, the trap is completely electrostatic, so there is no
limit on the mass that can be trapped. Second, the
trapping depends only on the ratio ${E_k}/{q}$, which means
that ions of different mass that are accelerated through the same potential difference can
be stored simultaneously. Third, the central part of the ion trap is (nearly)
field-free, so the ions travel in straight lines in this
region.

Various electrode voltage configurations are possible to achieve
trapping. We define a particular configuration by the set of
potentials $\{V_1,V_2,V_3,V_4,V_z\}$. $V_z$ is connected  to the
central electrode of an Einzel lens that plays the major role in
determining the focusing properties of the mirrors. In the present
work, only symmetric configurations, i.e., where identical
potentials are applied to the two mirrors, are considered. The
potentials on the four outmost electrodes were set to
$\{V_1,V_2,V_3,V_4\}$=\{6.5, 4.875, 3.25, 1.65\} kV, while the
Einzel voltage was varied between $2700<V_z<3200$ V and
$4000<V_z<4300$ V. These two ranges correspond to the known values
where the trap is stable, i.e., they satisfy the criterion
Eq.~\ref{eq:stable}, as has been shown in Ref.~\cite{Pedersen1}.
Additional details about trapping stability and the comparison to
optical models can be found in the literature\cite{Pedersen1}.

\subsection{Detection system}

One of the ion loss processes from the trap is charge
exchange, which leads to neutralization of the stored
particles. These neutral particles pass freely through the mirrors and can be
detected by a microchannel plate
(MCP) detector located downstream of the trap (see
Fig. \ref{expsetup}). The detector is coupled to a phosphor
screen so that the spatial
distribution of the neutral particles exiting the trap can be imaged. The
location and size of the MCP was different for the two
different ion source setups used in this work:  The MCP
was 25 mm in diameter, and located at a distance of 80.3 cm from
the center of the trap for the MALDI setup, while for the electron
impact ionization source the MCP was 40 mm in diameter, and
located at a distance of 90.3 cm from the center of the trap. The
imaging is performed by  a charge-coupled device (CCD) camera
located outside the vacuum that is connected to a frame
grabber which digitizes the picture in real time. The first image
is taken in coincidence with the raising of the potentials on the entrance mirror, and subsequent
images are digitized at a rate of 25 Hz for the whole trapping
time ($\sim1$ s). The positions of impact (($x,y$) on the front surface of the MCP)
are determined for all hits producing an amount of
light (as measured by the CCD camera) above a preprogrammed
threshold. Images of about 50 to 150 injections are averaged to
produce statistically significant results. The radial coordinate
$r$ is calculated as

\begin{equation}
r=\sqrt{(x-x_0)^2+(y-y_0)^2} \label{eq:radial}
\end{equation}

where ($x_0, y_0$) is the point where the trap axis crosses the
detector plane. This point is determined at a later stage by
finding the center of the measured radial distribution.

\subsection{Data analysis}

In order to study the TVD inside the trap, we analyze the radial
distribution of the neutral particles hitting the MCP detector.
Fig.~\ref{expsetup} shows the relationship between the ion
position and velocity inside the trap at the instant of its
neutralization, and the point of impact of the neutralized
particle on the detector, $r$. The ion's position at the
neutralization point is given by its distance $R$ from the optical
axis of the trap and distance from the MCP, $s$. The ion's
velocity at the same point is defined in terms of its longitudinal
and transversal velocities $v_\shortparallel$ and $v_\bot$,
respectively (see Fig.~\ref{expsetup}). If we assume that the
angular scattering taking place during the charge exchange is
small compared  to the angular dispersion of the beam (a very good
approximation for the heavy ions created in the MALDI
source)\cite{scattering}, then the position of impact on the MCP
can be calculated from

\begin{equation}
r=R+\frac{s v_{\bot}}{v_{\shortparallel}}.
\end{equation}

If we also use the fact that $R$ $\ll$ $r$, then we obtain for the
transverse velocity

\begin{equation}
v_{\bot}=\frac{r {v_{\shortparallel}}}{s}\approx
\frac{r}{s}\sqrt{\frac{2E_k}{m}}, \label{eq:vtrans}
\end{equation}

where $m$ is the particle mass.  Two problems arise from this
simple formula: First, the velocities $v_{\shortparallel}$ and
$v_{\bot}$ are not constant in the trap, as the particles are
slowed down and focused (or defocused, see Ref.~\cite{Pedersen1})
inside the mirrors. Second, the exact distance $s$ between the
neutralization point in the trap and the MCP is unknown. The
importance of these two effects, which can smear the radial
distribution measurement, will be treated separately using
numerical simulation, as described in the next Section.

\section{Numerical simulations}

In order to verify the different approximations made in the
derivation of Eq. \ref{eq:vtrans}, and to provide a better
understanding of the trap behavior, we have performed numerical
simulations of the particle trajectories in the actual potentials of
the ion trap. The calculations were carried out using SIMION
\cite{SIMION}, which can solve the Laplace equation for a specific
potential configuration in space and propagate ions on the
computed potential grid. The program uses a fourth-order
Runge-Kutta method to solve the Newtonian equations of motion. The
density of ions in the trap is assumed to be low enough for
ion-ion interactions to be neglected, and the trajectories are
calculated for one ion at a time (the actual number of ions in the
trap was of the order of 10$^5$ ions per injection).

For different values of $V_z$, while keeping the other potentials
constant, we have traced the stable trajectories, starting from an
initial distribution that covers the whole transverse stable
(i.e., trapped) phase space of the electrostatic trap, as
described in Ref.~\cite{Pedersen1}. The stable phase space was
found by systematically varying the initial conditions of the
particles. A stable trajectory was defined as one for which a
propagated ion was trapped for more than 500 $\mu$s (about 200
oscillations for 4.2 keV Ar$^+$, or 90 for 4.2 keV Au$^+$). It was
found that ions in unstable trajectories were usually lost from
the trap after a few oscillations ($<20\;\mu$s). The calculations
were made using a constant integration time step, and the
positions and velocities of the ions were recorded in a file at
each of these time steps. Using this information, simulated
distributions for the radial distribution at the MCP were
calculated by assuming an equal probability for neutralization at
each of these integration time steps, and propagating the
(neutral) particles in straight lines, using the initial positions
and velocities as recorded. This method has the advantage of
representing faithfully the local ion density along the length of
the trap. Implicit in the assumption of equal probability of
neutralization in each time step is the assumption that the
neutralization cross section is independent of kinetic energy for
energies below 4.2 keV\cite{neutralization}, even in the mirrors
where the kinetic energies approach zero. The results can then be
directly compared to the experimental distributions.

\section{Experimental and Simulation Results}

\begin{figure}[h]
\begin{center}
\includegraphics[width=15 cm]{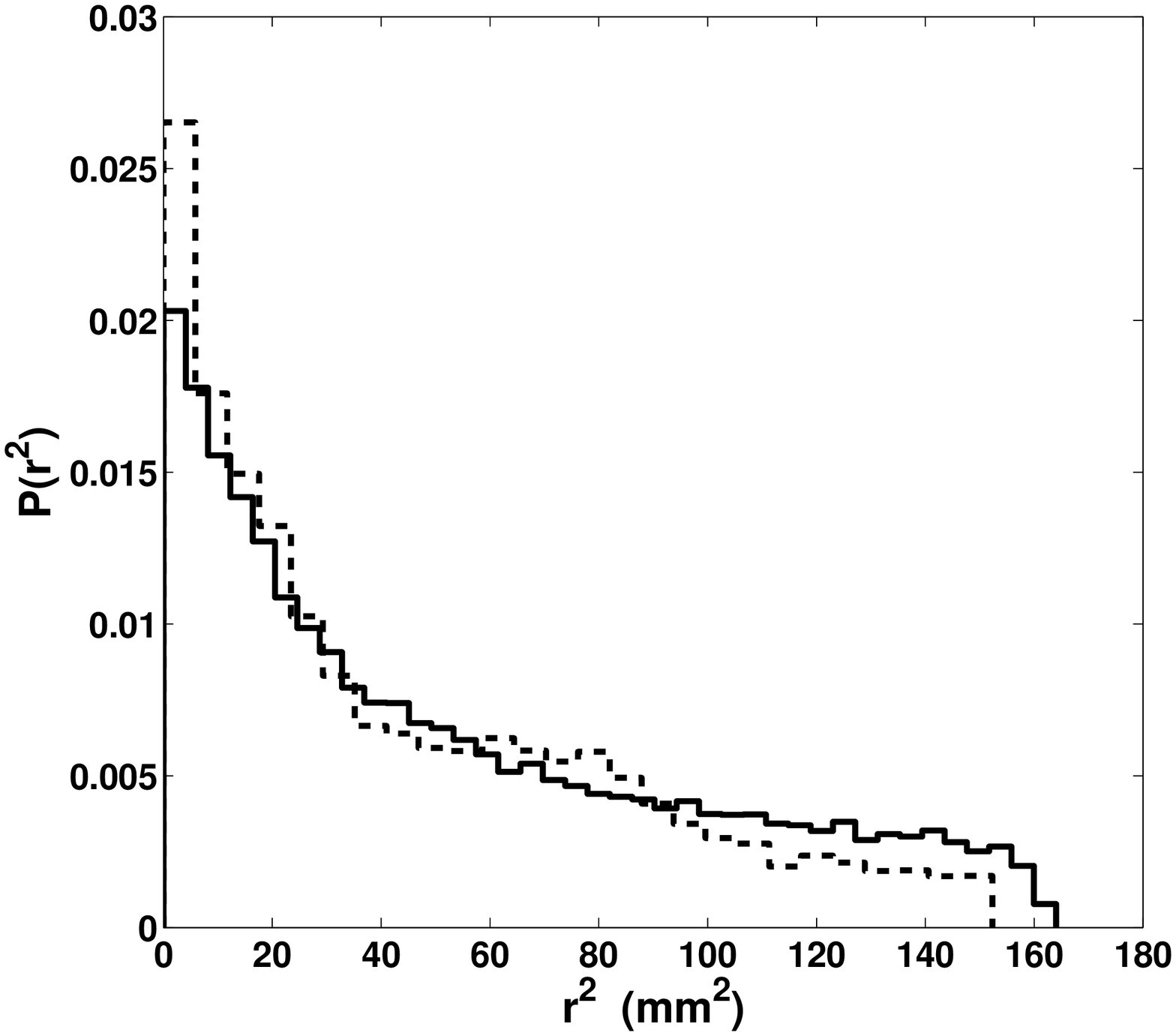}
\end{center}
\caption{}
\label{fig:r2hist}
\end{figure}

\begin{figure}[h]
\begin{center}
\includegraphics[width=14 cm]{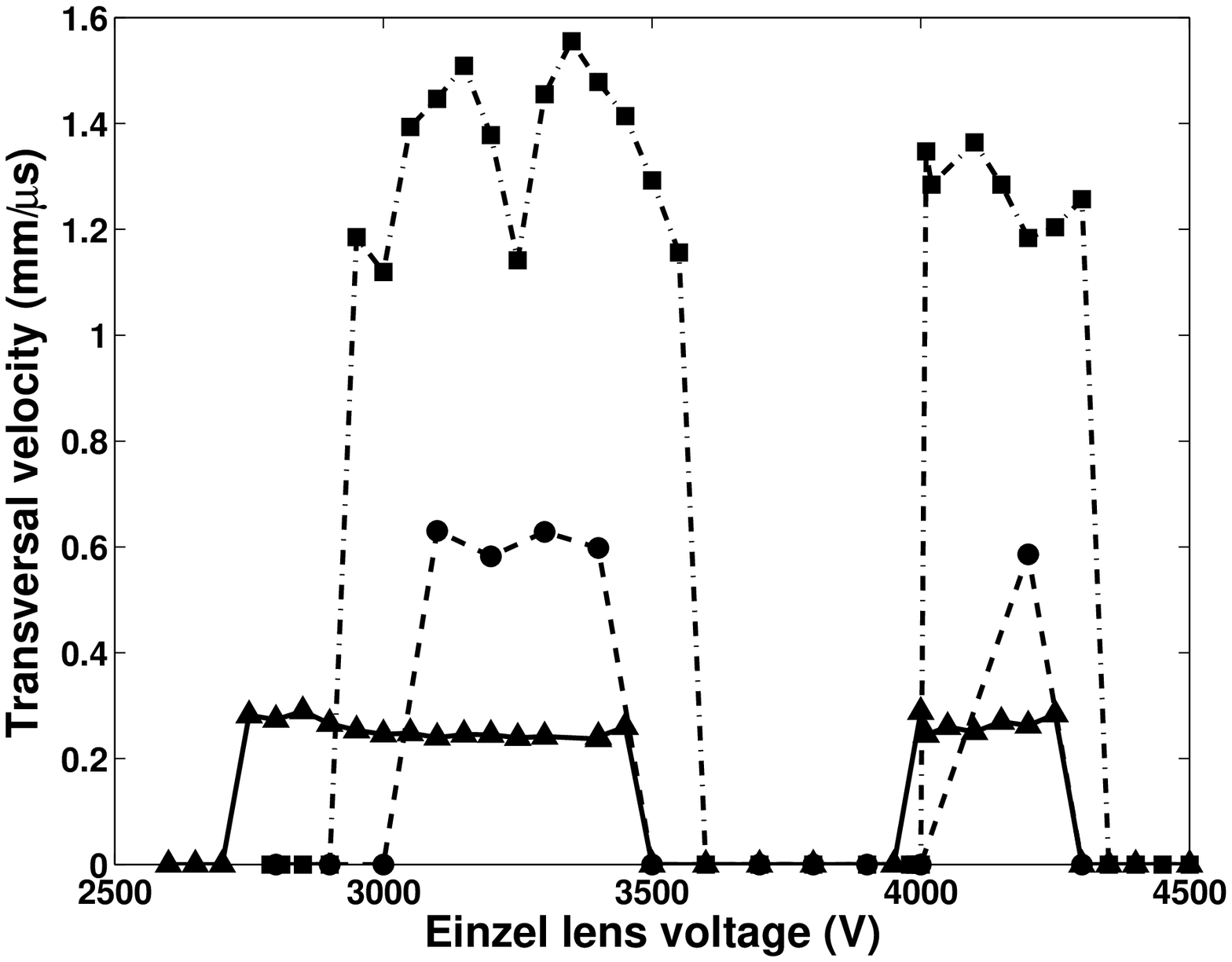}
\end{center}
\caption{}
\label{fig:vtransexp}
\end{figure}

Figure \ref{fig:r2hist} shows a comparison between the measured
(dotted line) and simulated (solid line) normalized distributions
for the distance squared ($P(r^2)$) at the MCP for 4.2 keV Au$^+$,
and $V_z$=3200 V. We choose to plot the $r^{2}$ distributions as
they display the radial density information in the most relevant
manner. The number of particles located in an interval of width
$d(r^2)=2rdr$ is proportional to the number of particles in the
ring between $r$ and $r+dr$, whose area is given by $2\pi rdr$.
Similar distributions were measured for Ar$^+$ and angiotensin
II$^+$ ions. Each of the measured distributions was characterized
by the standard deviation of the radial distribution which in the
present case is equal to the square root of the mean of r$^2$:
$\sigma_r=\sqrt{<r^2>}$. Using Eq.~\ref{eq:vtrans}, and replacing
$r$ by $\sigma_r$ and $s$ by the distance from the center of the
trap to the MCP, typical transverse velocities $v_{\bot}$ could be
obtained for the different masses and values of the  Einzel lens voltage ($V_z$).
Figure ~\ref{fig:vtransexp} shows the results for the three different
ions as a function of $V_z$. Only a weak
dependence of transverse velocity $v_\bot$ on $V_z$ is observed,
except for Ar$^+$ around 3250 V. The ratio
$v_{\bot}/v_{\shortparallel}\approx$ 9$\times$10$^{-3}$ is found
to be approximately constant for all ions.

\begin{figure}[h]
\begin{center}
\includegraphics[width=12 cm]{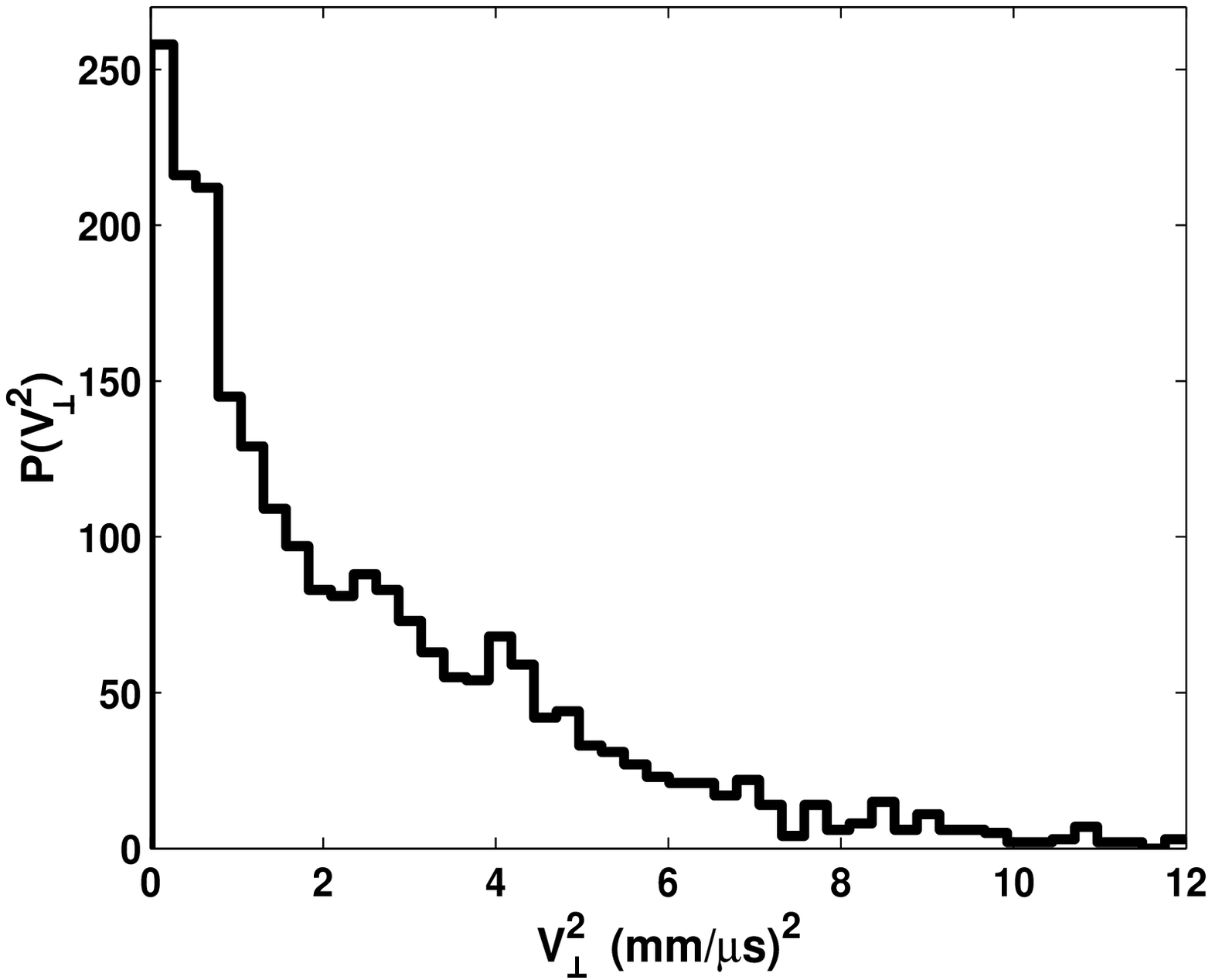}
\end{center}
\caption{}
\label{fig:vtranssim}
\end{figure}

Based on the excellent agreement between the experimental data and
the simulations (see Fig.~\ref{fig:r2hist}), one can now use the
simulation to check the assumptions which led to
Eq.~\ref{eq:vtrans}, especially the assumption related to the
contribution of the neutral particles produced inside the mirrors
to $P(r^2)$ and the unknown distance between the neutralization
point and the MCP. Figure~\ref{fig:vtranssim} shows an example of
the distribution of the square of the transverse velocity
$P(v_{\bot}^2)$ from Simion simulations. The case presented is for
all stable Ar$^+$ ions in the field-free region of the trap, for
an Einzel lens voltage of $V_z$=3300 V. To compare this
distribution to the experimentally deduced typical transverse
velocity, we characterize this distribution in a similar way as
the squared radial distribution (see Fig.~\ref{fig:r2hist}), using
the square root of its mean, which is equivalent to the standard
deviation of the TVD, yielding $\sigma_{v{\bot}}$=1.16 mm/$\mu$s.
This value is slightly lower than the one derived directly from
the measured radial distributions (see Fig.~\ref{fig:vtransexp}),
as can be expected since the latter includes some contributions
from slower ions inside the mirrors that tend to have also larger
$v_\bot/v_\shortparallel$ ratio. However, the difference is
relatively small (the reduced detection efficiency of the MCP for
the slower particles also contributes to the fact that these have
a minor effect on the measured distributions), and we conclude
that the data shown in Fig.~\ref{fig:vtransexp} are an upper limit
of the transverse velocity of the trapped ions in the central
(field-free) region of the trap.

\section{Transverse phase space}

The results obtained in the previous section show that the
measured and simulated transverse velocities are in good
agreement. Since the simulated value is obtained by filling the
stable phase space of the trap, one can conclude that the ions
stored in the trap always fill the available (stable)
transverse phase space. This has an important implication as far
as the ion loss processes are concerned. As pointed out
previously \cite{Pedersen1}, two processes play an important
role in limiting the lifetime of the ions in the trap. The first
is neutralization of the ions via charge exchange
with the residual gas molecules, and the second is
multiple scattering, which increases the transverse velocity of the
ions until they reach the limit of the stable transverse phase
space. Although the importance of the neutralization process can
be observed experimentally by counting the number of neutral
particles exiting the trap, the importance of multiple
scattering is more difficult to observe experimentally. Also, it
is quite clear that the loss due to neutralization is independent of
the trap configuration, while the loss due to multiple scattering
will be strongly dependent on the available transverse phase
space, if the stable transverse phase space is always full. The
fact that the measured transverse velocity is found to be equal to
the one extracted from a simulated "full transverse phase space"
suggests that the loss due to multiple scattering is very
important, and that the lifetime of the ions is mostly limited by
this process, a conclusion which was already reached by
Pedersen et al. \cite{Pedersen1} using arguments based on known angular
scattering cross sections.

\begin{figure}[h]
\begin{center}
\includegraphics[width=12 cm]{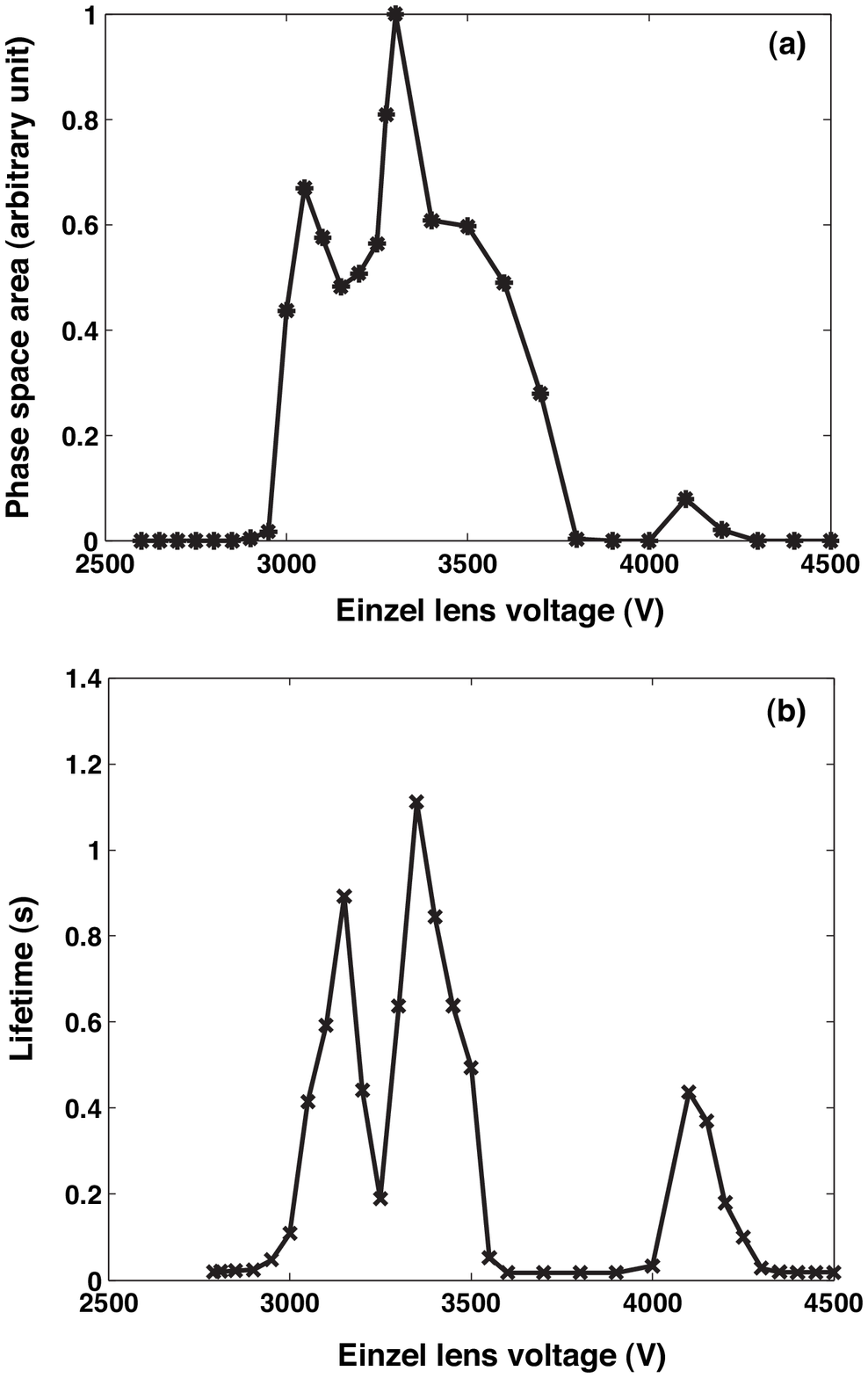}
\end{center}
\caption{}
\label{fig:lifetime}
\end{figure}

In order to demonstrate the importance of multiple scattering as a
loss process, we have calculated the area of the stable transverse
phase space for the Ar$^+$ ions as a function of the Einzel lens
voltage $V_z$. The area was calculated by recording the transverse
position and velocity for each pass of a stable ion through the
midplane of the trap. A scatter plot of these coordinates was then
made, and the area filled by the points was estimated by dividing
the phase space into a fine grid and counting the number of cells
for which at least four points were found. These cells are then
called "stable cells" (the minimum number of points required for a
cell to be defined as stable has only a small influence on the
final results). Fig.~\ref{fig:lifetime}(a) shows the results of
such a calculation as a function of V$_z$, while
Fig.~\ref{fig:lifetime}(b) shows the lifetime of the ions in the
trap, as obtained by measuring the rate of neutral Ar
hitting the MCP as a function of storage time and fitting the
decay using an exponential function. A clear correspondence
between the lifetime and the area of the transverse phase space is
observed, including the dip around V$_z$=3200 V.

\section{Conclusions}
We have measured the transverse velocity distribution of 4.2 keV
Ar$^+$, Au$^+$, and angiotensin II$^+$ stored in a linear
electrostatic ion trap. The results show that the width of the TVD
is mass dependent, and represents about 1\% of the longitudinal
beam velocity for the present trap geometry. The experimental
results are in excellent agreement with the numerical simulation.
More important, it shows that in the existing experimental setup,
the phase space of the trap is filled very soon after injection.
Thus, we can expect multiple scattering to be an important ion
loss process (the other being neutralization). This is also
demonstrated by the correlation between the area of transverse
phase space and the measured lifetimes.

A consequence of our results is that the lifetime in the trap will
also be a function of the trap length. Indeed, for a given angular
dispersion of the beam and for a given mirror geometry, a trap
with longer distance between the mirrors will be less stable, as
the particles will be further away from the central axis of the
trap when they enter the mirrors. On the other hand, although
shorter traps will probably be more stable, they can store less
ions.

The tool that we have developed to measure the transverse velocity
distribution of the stored ions can now be used in studies of
transverse cooling. Specifically, if a kicker for stochastic
cooling is installed in the field-free region of the trap, it
should be possible to shrink the $r^2$ distribution, and thereby
increase the storage lifetime. Moreover, because of Coulomb
repulsion between the ions (especially near the turning points in
the mirrors) and the radial mixing induced by the mirror, we can
expect that transverse cooling will also affect the longitudinal
velocity distribution.

This work was supported in part by the Israel Science Foundation.
Laboratoire Kastler Brossel is Unit{\'e} Mixte de Recherche du CNRS no. 8552,
of the Physics D{\'e}partement of Ecole Normale Sup{\'e}rieure and Universit{\'e} Pierre
et Marie Curie.

\newpage

{\bf Figures}

\newpage

\newpage

\newpage

\newpage

\newpage

\clearpage

{\bf Figure Captions}

\begin{itemize}

\item Fig. 1: Schematic view of the experimental setup. The bunch
is injected through the left hand side of the trap (entrance
electrodes), and the neutral particles escaping from the trap
through the exit electrodes are counted by the MCP detector, whose
phosphor screen is imaged by a CCD camera. The drawing is not to
scale.

\item Fig. 2: Normalized $r^{2}$ distributions for 4.2 keV Au$^+$.
Dotted line: measured by the MCP detector; solid line: simulation
with Simion\cite{SIMION}. The Einzel lens voltage was 3200 V.

\item Fig. 3: Measured transverse velocity as a function of the
Einzel lens voltage for three different ions: Ar$^+$
($\blacksquare$); Au$^+$ ($\bullet$); angiotensin II$^+$
($\blacktriangle$). The transverse velocity is set to zero for
unstable configurations.

\item Fig. 4: Calculated transverse velocity distribution in the
field-free region of the trap for Ar$^+$ ions with $V_z$=3300 V.

\item Fig. 5: (a) Area of the stable transverse phase space in the
midplane of the trap as a function of $V_z$. (b) Measured lifetime
for Ar$^+$ as a function of $V_z$.

\end{itemize}


\begin{thebibliography}{10}
\bibitem{cite1}
  W.Paul and H.Steinwedel, Z. Naturforsch A {\bf 8} 448 (1953).

\bibitem{ionTrapping}
  For a review on ion trapping, see Phys. Scripta. {\bf T59}
  (1995), and {\it proceedings of the Conference on Trapped Charged Particles and Fundamental
  Interactions}, J. Phys. B {\bf 36}, issue 3 and 5 (2002).

\bibitem{EIBT1}
  D. Zajfman et al., Phys. Rev. A {\bf 55}, 1577 (1997).

\bibitem{EIBT2}
  M. Dahan et al., Rev. Sci. Instrum. {\bf 69}, 76 (1998).

\bibitem{meta1}
R. Wester, K.G. Bhushan, N. Alstein, D. Zajfman, O. Heber, and
M.L. Rappaport, J. Chem. Phys. {\bf 110}, 11 830 (1999).

\bibitem{meta2}
K.G. Bushan, H.B. Pedersen, N. Alstein, O. Heber, M.L. Rappaport,
and D. Zajfman, Phys. Rev. A {\bf 62}, 012504 (2000).

\bibitem{life1}
A. Wolf, K.G. Bhushan, I. Ben-Itzhak, N. Alstein, D. Zajfamn, O.
Heber and M.L. Rappaport, Phys. Rev. A {\bf 59}, 267 (1999).

\bibitem{life2}
L. Knoll, K.G. Bhushan, N. Alstein, D. Zajfman, O. Heber and M.L.
Rappaport Phys. Rev. A {\bf 60}, 1710 (1999).

\bibitem{adi}
A. Diner et al., Phys. Rev. Lett. {\bf 93}, 063402 (2004).

\bibitem{yoni}
A. Naaman et al., J. Chem. Phys., 113 4662 (2000)


\bibitem{bunch1}
  H.B. Pedersen, D. Strasser,S. Ring, O. Heber, M.L. Rappaport, Y
  Rudich, I. Sagi and D. Zajfman, Phys. Rev. Lett. {\bf 87} 055001-1
  (2001).

\bibitem{sarah} S. Goldberg, D. Strasser, O. Heber, M. L. Rappaport,
  A. Diner, and D. Zajfman, Phys. Rev. A {\bf 68}, 043410 (2003).

\bibitem{Moller:97}
S. P. M\o ller,  Nucl. Instrum. Methods in Physics Research A {\bf
394}, 281 (1997).

\bibitem{Moller:01}
S. P. M\o ller and U. V. Pedersen, Phys. Scripta  {\bf T92}, 105
(2001).

\bibitem{He-:01}
U. V. Pedersen, M. Hyde, S. P. M\o ller and T. Andersen,  Phys.
Rev. A {\bf 64}, 012503-1 (2001)

\bibitem{Hansen:01}
K. Hansen, J. U. Andersen, P. Hvelplund, S. P. M\o ller, U. V.
Pedersen, and V. V. Petrunin, Phys. Rev. Lett. {\bf 87}, 123401
(2001).

\bibitem{maldi}
F. Hillenkamp et al., Anal. Chem {\bf63}, 1193A (1991)

\bibitem{Pedersen1}
  H.B. Pedersen, D. Strasser, O. Heber, M.L. Rappaport and D. Zajfman,
  Phys. Rev. A {\bf65} 042703 (2002).

\bibitem{scattering}
L. K. Johnson, R. S. Gao, C. L. Hakes, K. A. Smith, and R. F.
Stebbings, Phys. Rev. A {\bf 40}, 4920 (1989).

\bibitem{SIMION}
SIMION, Version 6.0, Ion Source Software \\
(http://www.srv.net/$\sim$klack/simion.html).

\bibitem{neutralization}
G. J. Lockwood, Phys. Rev. A {\bf 2}, 1406 (1970).



\end{thebibliography}
\end{document}